\documentclass[a4paper,12pt]{article}
\usepackage{amssymb,amsmath}
\usepackage{epsfig,fancyheadings}

\usepackage{a4}
\usepackage{parskip}
\usepackage{color,soul}
\usepackage{amsthm,amsxtra,amscd,relsize,multirow}
\usepackage[all,2cell]{xy}\UseAllTwocells

\newcommand{\sect}[1]{\setcounter{equation}{0}\section{#1}}
\renewcommand{\theequation}{\arabic{section}.\arabic{equation}}

\def\ds{\displaystyle}

\def\a{\alpha}
\def\b{\beta}
\def\d{\delta}

\def\x{\xi}
\def\z{\zeta}
\def\m{\mu}

\def\f{\phi}
\def\vf{\varphi}
\def\e{\eta}

\def\o{\omega}
\def\sinh{\mathrm{sinh}}
\def\cosh{\mathrm{cosh}}

\def\p{\partial}
\def\rb{\right}
\def\lb{\left}
\def\axs{AdS_5\times S^5}
\newcommand{\eq}[1]{\begin{equation} #1 \end{equation}}
\newcommand{\al}[1]{\begin{align} #1 \end{align}}
\newcommand{\ml}[1]{\begin{multline} #1 \end{multline}}

\def\cp{\mathbb {CP}^3}

\begin{titlepage}
\title{A note on the reduction of the $AdS_4\times\cp$ string $\sigma$-model}

\author{R.~C.~Rashkov${}^{\dagger}$\footnote{e-mail:
rash@hep.itp.tuwien.ac.at; also Dept of Physics, Sofia University,
Bulgaria.}
\ \\ \ \\
${}^{\dagger}$ Institute for Theoretical Physics, \\ Vienna
University of Technology,\\
Wiedner Hauptstr. 8-10, 1040 Vienna, Austria }
\date{}
 \end{titlepage}

\begin{document}

\maketitle
\thispagestyle{fancy}

 \begin{abstract}
We study the reduction of the (bosonic) string sigma model on
$AdS_4\times\cp$ background. We give a brief review of the known
results for the $AdS$ part and apply an explicit reduction scheme
to the $\cp$ part of the model. A brief discussion on the reduced
model is presented.
  \end{abstract}

\section{Introduction}

 The correspondence between the large N limit
of gauge theories and string theory was on the focus of intensive
promising research for more than thirty years and in different
periods it showed different faces. One of the most promising
explicit realizations of this correspondence was provided by the
Maldacena conjecture about AdS/CFT correspondence
\cite{Maldacena}. Being an excellent example of exact duality
between type IIB string theory on $AdS_5\times S^5$ and
$\mathcal{N}=4$ super Yang-Mills theory
\cite{Maldacena,GKP,Witten}, this subject has became a major
research and many fascinating developments have been done.

A cornerstone in the current understanding of the duality between
gauge theories and strings (M-theory) is the world-volume dynamics
of the branes. Recently there has been a considerable amount of
work focused on the understanding of the worldvolume dynamics of
multiple M2-branes - an interest inspired by Bagger, Lambert and
Gustavsson \cite{Bagger:2006sk} and their investigations based on
the structure of Lie 3-algebra.

Recently, inspired by the study of the Bagger-Lambert-Gustavson
theory on $N$ membranes and motivated by the possible description
of the worldvolume dynamics of coincident membranes in M-theory, a
new class of conformal invariant, maximally supersymmetric field
theories in 2+1 dimensions has been found
\cite{Schwarz:2004yj,ABJM}. The main feature of these theories is
that they contain gauge fields with Chern-Simons  kinetic terms.
Based on this development, Aharony, Bergman, Jafferis and
Maldacena proposed a new gauge/string duality between an
$\mathcal{N}=6$ super-conformal Chern-Simons theory (ABJM theory)
coupled to bi-fundamental matter, describing $N$ M2 brnaes on
$\mathbb{R}^8/\mathbb{Z}_k$. This model is believed to be dual to
M-theory on $AdS_4\times S^7/\mathbb Z_k$.

The ABJM theory actually has two Chern-Simons gauge fields with
opposite levels, $k$ and $-k$ correspondingly, each with gauge
group $SU(N)$ (or $U(N)$). The two pairs of chiral superfields
transform in the bi-fundamental representations of $SU(N) \times
SU(N)$ and the R-symmetry is $SU(4)$ as it should be for $\mathcal
N=6$ supersymmetry of the theory. It was observed in \cite{ABJM}
that there exists a natural definition of a 't Hooft coupling --
$\lambda = N/k$. Ii was observed that in the 't Hooft limit
$N\rightarrow \infty$ with $\lambda$ held fixed, one has a
continuous coupling  $\lambda$ and the ABJM theory is weakly
coupled for $\lambda \ll 1$. The ABJM theory is conjectured to be
dual to M-theory on $AdS_4\times S^7 / \mathbb Z_k$ with $N$ units
of four-form flux. In the scaling limit $N, k \rightarrow\infty$
with $k \ll N \ll k^5$ the theory reduces to type IIA string
theory on $AdS_4 \times \mathbb{CP}^3$. Thus, the AdS/CFT
correspondence, which has led to many exciting developments in the
duality between type IIB string theory on $AdS_5\times S^5$ and
${\cal N}=4$ super Yang-Mills theory, is now being extended to the
$AdS_4/CFT_3$ and is expected to constitute a new example of exact
gauge/string theory duality.

Semi-classical strings have played, and still play, an important
role in studying various aspects of $AdS_5/SYM_4$ correspondence
\cite{Bena:2003wd}-\cite{Lee:2008sk}. The development in this
subject gived a strong hint about how the new emergent
$AdS_4/CFT_3$ duality can be investigated. An important role in
these studies is played by integrability. The superstrings on
$AdS_4\times\cp$ as a coset was first studied in
\cite{Arutyunov:2008if}\footnote{See also \cite{Stef}} which opens
the door for investigation of the integrable structures in the
theory. Various properties on the gauge theory side and tests on
string theory side as rigid rotating strngs, pp-wave limit,
relation to spin chains, as well as pure spinor formulation  have
been considered \cite{Arutyunov:2008if}-\cite{D'Auria:2008cw}. In
these intensive studies many properties were uncovered and
impressive results obtained, but still the understanding of this
duality is far from complete.


\paragraph{ABJM and strings on $AdS_4\times \cp$}

To find the ABJM theory one starts with analysing M2-brane dynamics governed by eleven-dimensional 
supergravity action \cite{ABJM} 
\eq{ 
S=\frac{1}{2\kappa_{11}^2}\int
dx^{11}\sqrt{-g}\left(R-\ds\frac{1}{2\cdot
4!}F_{\mu\nu\rho\sigma}F^{\mu\nu\rho\sigma}\right)-\frac{1}{12\kappa_{11}^2}
\int C^{(3)}\wedge F^{(4)}\wedge F^{(4)}, \label{abjm-1} 
}
 where $\kappa_{11}^2=2^7\pi^8 l_p^9$. Solving for the equations of motions
\eq{
R^\mu_\nu=\frac{1}{2}\left(\frac{1}{3!}F^{\mu\a\beta\gamma}F_{\nu\a\beta\gamma}
-\frac{1}{3\cdot 4!}\delta^\mu_\nu
F_{\a\beta\rho\sigma}F^{\a\beta\rho\sigma}\right), \label{abjm-2}
} 
and 
\eq{
\p_{\sigma}(\sqrt{-g}F^{\sigma\mu\nu\xi})=\frac{1}{2\cdot
(4!)^2}\epsilon^{\mu\nu\xi\a_1\dots\a_8}F_{\a_1\dots\a_4}F_{\a_5\dots\a_8},
\label{abjm-3} 
} 
one can find the M2-brane solutions whose near horizon limit becomes $AdS_4\times S^7$ 
\eq{
ds^2=\frac{R^2}{4}ds^2_{AdS_4}+R^2 ds^2_{S^7}. \label{abjm-4} 
} 
In addition we have $N'$ units of four-form flux 
\eq{
 F^{(4)}=\frac{3R^3}{8}\epsilon_{AdS_4}, \quad R=l_p(2^5 N'\pi^2)^{\frac{1}{6}}.
\label{abjm-5} 
} 
Now one proceeds with considering the quotient $S^7/\mathbb Z_k$ acting as 
$z_i \rightarrow e^{ i\frac{ 2 \pi}{ k} } z_i$. It is convenient first to write the metric on $S^7$ as
\eq{ 
ds^2_{S^7} =  ( d \varphi' + \omega)^2 + ds^2_{CP^3},
\label{abjm-6} 
} 
where 
\al{
 & ds^2_{CP^3} = \frac{ \sum_i d z_i d
\bar z_i}{ r^2 } -
    \frac{ | \sum_i z_i d \bar z_i |^2}{ r^4} ~,\quad
 r^2 \equiv \sum_{i=1}^4 |z_i|^2,\notag \\
& d \varphi' + \omega  \equiv  \frac{ i}{ 2  r^2 }\sum\limits_i (
z_i d \bar z_i - \bar z_i d z _i ),\quad d\omega =  J = {
i}\sum\limits_i   d \left(\frac{ z_i }{r}\right)  d \left(\frac{
\bar z_i}{r }\right). \label{abjm-7} 
} 
and then to perform the $\mathbb Z_k$ quotient identifying $\vf'=\vf/k$ with $\vf\sim
\vf+2\pi$ ($J$ is proportional to the K\"{a}hler form on $\cp$). The resulting metric becomes 
\eq{ 
ds^2_{S^7/{\mathbb Z}_k} =
\frac{ 1 }{k^2 } ( d \vf+ k \o)^2 + ds^2_{CP^3}. 
} 
One observes that the first volume factor on the right hand side is divided by
factor of $k$ compared to the initial one. In order to have consistent quantized flux one must impose 
$N'=kN$ where $N$ is the number of quanta of the flux on the quotient. One should note that
the spectrum of the supergravity fields of the final theory is just the projection of the initial 
$AdS_4\times S^7$ onto the $\mathbb Z_k$ invariant states. In this setup there is a natural
definition of {}'t Hooft coupling $\lambda \equiv N/k$. Decoupling limit should be taken as $N,k\rightarrow \infty$ 
while $N/k$ is kept fixed.

One can follow now \cite{ABJM} to make reduction to type IIA with
the following final result 
\al{
 ds^2_{string} = & \frac{ R^3}{ k} ( \frac{ 1 }{4 } ds^2_{AdS_4} + ds^2_{{\mathbb CP}^3 } ),\\
 e^{2 \phi} = & \frac{ R^3}{ k^3 } \sim \frac{ N^{1/2}}{ k^{5/2} }= \frac{ 1 }{ N^2 } \left(
\frac{ N }{k } \right)^{5/2}, \\
 F_{4} = & \frac{ 3}{ 8 }  {  R^3}  \epsilon_4 ,
 \quad F_2 =  k d \omega = k J,
 }
 We end up then with  $AdS_4 \times\cp$ compactification  of type IIA string theory with $N$ units of 
$F_4$ flux on $AdS_4$ and $k$ units of $F_2$ flux on the ${\mathbb CP}^1 \subset\cp$ 2-cycle.

The radius of curvature in string units is $R^2_{str} = \frac{R^3}{ k}  =  2^{5/2} \pi \sqrt{ \lambda}$. 
It is important to note that the type IIA approximation is valid in the regime where 
$k \ll N \ll k^5$.


To fix the notations, we write down the explicit form of the
metric on $AdS_4\times\mathbb{CP}^3$ in spherical coordinates. The
metric on $AdS_4\times\mathbb{CP}^3$ can be written as \cite{PopeWarner}
\begin{multline}
 ds^2=R^2\left\lbrace \dfrac{1}{4}\left[ -\cosh^2\rho\,dt^2+d\rho^2+\sinh^2\rho\,d\Omega_2^2\right] \right.\\
\left.+d\mu^2+\sin^2\mu\left[d\alpha^2+\dfrac{1}{4}\sin^2\alpha(\sigma_1^2
+\sigma_2^2+\cos^2\alpha\sigma_3^2)+\dfrac{1}{4}\cos^2\mu(d\chi+\sin^2\mu\sigma_3)^2\right]\right\rbrace.
\label{metric}
\end{multline}
Here $R$ is the radius of the $AdS_4$, and $\sigma_{1,2,3}$ are
the $SU(2)$ left-invariant 1-forms, parameterized by
$(\theta,\phi,\psi)$,
\begin{align}
&\sigma_1=\cos\psi\,d\theta+\sin\psi\sin\theta\,d\phi,\notag\\
&\sigma_2=\sin\psi\,d\theta-\cos\psi\sin\theta\,d\phi,\label{S3}\\
&\sigma_3=d\psi+\cos\theta\,d\phi.\notag
\end{align}
The range of the coordinates is
$$0\leq\mu,\,\alpha\leq\dfrac{\pi}{2},\,\,0\leq\theta\leq\pi,\,\,0\leq\phi\leq2\pi,\,
\,0\leq\chi,\,\psi\leq4\pi.$$


\sect{Reduction of $AdS_4\times\cp$ sigma models}

\subsection{The reduction of $AdS_4$ }

In this Section we present the reduction of the $AdS_4$ part of
the string sigma model. Although most of the results in this
section are known, it would be useful to review some methods of
reduction which were proved to be useful in the context of
strings. First we consider a string moving only in the $AdS_4$
part of spacetime and then assume that the motion is not
constrained to that part. The difference is as follows.
Considering the dynamics of the string on $AdS_4\times\cp$ we will
allways assume that the Virasoro constraints are satisfied, i.e.
\eq{ 
T_{\pm\pm}^{AdS}+T_{\pm\pm}^{\cp}=0, \quad
T_{\pm\pm}^{AdS}=-\kappa^2, \quad T_{\pm\pm}^{\cp}=\kappa^2.
\label{Vir-tot} 
} 
When the dynamics is confined only to the $AdS_4$ part of the geometry $T_{\pm\pm}^{\cp}=0$ and thus, one
can distinguish two cases. We present below both of them using slightly different approaches.

\paragraph{Reduction of pure $AdS_4$.}\ \\

To apply the reduction scheme one must represent the $AdS_n$ space
as a coset space $SO(n,1)/SO(n-1,1)$. Then the string sigma model
can be thought of as sigma model on the above symmetric space.
Here we follow the approach developed in
\cite{Pohl,Vega,Jev-2,Grigoriev:2007bu,Grigoriev:2008jq}.

Let us restrict our attention to the specific case of $AdS_4$
spacetime and consider it as a hyperboloid embedded into
five-dimensional Euclidean space: \eq{ \vec Y.\vec Y\equiv
-Y_{-1}^2 - Y^2_0+Y^2_1+Y^2_2+Y^2_3=-1 \label{embedd-ads} }
Writing the action in these variables (plus the above constraint),
one finds the equations of motion 
\eq{ \vec Y_{\x\e}-(\vec
Y_\x.\vec Y_\e)\vec Y=0, \label{EOM-p}
} 
where $Y_\x\equiv \p_\x Y$ etc.\footnote{From now on the subscripts $\x$ and $\e$ denote
derivatives withe respect to the corresponding variable.}, as usual $\x=\frac{1}{2}(\sigma+\tau), \,\,
\e=\frac{1}{2}(\sigma-\tau)$.

The corresponding Virasoro constraints have the explicit form 
\eq{
\vec Y_\x.\vec Y_\x=\vec Y_\e.\vec Y_\e=0. \label{vir-p} 
} 
To explicitly carry out the reduction we introduce a basis 
\eq{
\{\vec e_i\}=\{\vec Y,\vec Y_\x, \vec Y_\e, \vec e_4, \vec e_5\}
\label{basis-p} 
} 
where for $i=1,2,3$ the properties of the basis vectors are dictated by the embedding \eqref{embedd-ads} and
equations of motion \eqref{EOM-p} while for $i=4,5$ we require the following conditions to be satisfied 
\al{
& \vec e_i{}^2=1, \notag  \\
& \vec e_i.\vec Y=\vec e_i.\vec Y_\x=\vec e_i.\vec Y_\e=0..
\label{def-bas-p} 
} 
The Virasoro constraints take the form 
\al{
& T_{\x\x}=\vec Y^2_\x=0 \notag \\
& T_{\e\e}=\vec Y^2_\e=0 \label{vir-p-1} 
} 
and $T_{\x\e}\equiv 0$ automatically (note also that $\vec Y_\x$ and $\vec Y_\e$ are
null-vectors).

Now we define the angle $\a(\x,\e)$ through (Liouville mode for
the case of $AdS_2$) 
\eq{ \vec Y_\x.\vec Y_\e=e^{\a(\x,\e)}.
\label{def-sinh} 
} 
One can find some useful relations from the above definitions 
\eq{
 \vec Y.\vec Y_\x=\vec Y.\vec Y_\e=0, \quad
 \vec Y_\x.\vec Y_{\x\x}=\vec Y_\e.\vec Y_{\e\e}=0, \quad
} 
Next we want to express the second derivatives expanded over the
basis \eqref{basis-p}. One of them follows immediately from the
definitions and the properties above 
\eq{ \vec
Y_{\x\e}=e^{\a(\x,\e)}\vec Y, \label{p1} 
} 
but we want also to find the other second derivatives of $\vec Y$. To find them we
expand over the basis $\{e_k\}$ \eqref{basis-p}\footnote{Since $\vec Y.\vec Y_{\x\x}=0$ there is 
no term proportional to $\vec Y$.} 
\al{
& \vec Y_{\x\x}=A\vec Y_\x+ B\vec Y_\e+\lb(a_4\vec e_4+a_5\vec e_5\rb) \label{p2-a} \\
& \vec Y_{\e\e}=C\vec Y_\x+ D\vec Y_\e+ \lb(b_4\vec e_4+b_5\vec
e_5\rb) \label{p2-b} 
} 
To obtain the coefficients $a_i$ one must multiply \eqref{p2-a} with $\vec e_i$ and take into account the
orthogonality conditions \eqref{def-bas-p} (analogously for $b_i$). The result is 
\eq{
 a_i=\vec Y_{\x\x}.\vec e_i, \quad b_i=\vec Y_{\e\e}.\vec e_i
} 
We want to obtain the coefficients $A,B,C,D$. Using the orthogonality of the basis and 
\eqref{def-sinh} one finds 
\eq{
B=C=0. 
} 
From 
\eq{ 
\vec Y_\e.\vec Y_{\x\x}=Ae^\a, \quad \vec
Y_\x.\vec Y_{\e\e}=De^\a \quad \text{and} \quad \a_\e e^\a=\vec
Y_\x.\vec Y_{\e\e} 
} 
we find 
\al{
& A=\a_\x(\x,\e) \\
& D=\a_\e(\x,\e) 
} 
The final form of the second derivatives of $\vec Y$ is 
\al{
& \vec Y_{\x\x}=\a_\x(\x,\e)\vec Y_\x+ \lb(a_4\vec e_4+a_5\vec e_5\rb)\label{p3-a}\\
& \vec Y_{\e\e}=\a_\e(\x,\e)\vec Y_\e+ \lb(b_4\vec e_4+b_5\vec
e_5\rb) \label{p3-b} 
} 
To obtain the equation for $\a(\x,\e)$ we must eliminate all the $\vec Y$ and its derivatives. First,
differentiating \eqref{def-sinh} with respect to $\e$ one finds (using also that 
$\vec Y_{\x\e}=e^\a \vec Y$ and $\vec Y.\vec
Y_\e=0$) 
\eq{ 
\a_\e(\x,\e)=e^{-\a(\x,\e)}\,\vec Y_\x.\vec Y_{\e\e}
} 
Differentiating the above equation with respect to $\x$ we get
\eq{ 
\a_{\x\e}(\x,\e)=e^{-\a}\lb[-\a_\x\vec Y_\x.\vec
Y_{\e\e}+\vec Y_{\x\x}.\vec Y_{\e\e}+\vec Y_\x.\vec Y_{\e\e\x}\rb]
\label{p5} 
} 
Combining the properties of the orthogonal basis from
\eqref{p5} we find 
\eq{
\a_{\x\e}(\x,\e)-e^{\a(\x,\e)}-e^{-\a(\x,\e)}\lb(a_4
b_4+a_5b_5\rb) =0 \label{p4-a} 
}

It is a simple exercise to cast the resulting equations in linear
form 
\al{
\frac{d}{d\x}\vec e_i(\x,\e)=A_{ij}(\x,\e)\vec e_j(\x,\e) \label{p5-a} \\
\frac{d}{d\e}\vec e_i(\x,\e)=B_{ij}(\x,\e)\vec e_j(\x,\e),
\label{p5-b} 
} 
with compatibility condition 
\eq{ 
\p_\e
\mathbf{A}-\p_\x\mathbf{B}+[\mathbf{A},\mathbf{B}]=0.
\label{compatib} 
} 
To obtain the entries of the matrices $\mathbf{A}$ and $\mathbf{B}$ one has to use the orthogonality
conditions and the properties discussed above. Skipping the details, for the matrix $\mathbf{A}$ we find
 \eq{ 
\mathbf{A}=
\begin{pmatrix}
 0      & 1      & 0           & 0      & 0    \\
 0      & \a_\x  & 0           & a_4    & a_5   \\
 e^\a   & 0      & 0           & 0      & 0       \\
 0      & 0      & -a_4e^{-\a} & 0      & (\vec e_{4\x}.\vec e_5)   \\
 0      & 0      & -a_5e^{-\a} &  (\vec e_{5\x}.\vec e_4)     & 0
\end{pmatrix},
\label{p8} 
} 
while for the other matrix, \textbf{B}, we find 
\eq{
\mathbf{B}=
\begin{pmatrix}
 0      & 0      & 1           & 0      & 0    \\
 e^\a   & 0      & 0           & 0      & 0   \\
 0      & 0      & \a_\e       & b_4    & b_5       \\
 0      & -b_4e^{-\a}      & 0 & 0      & (\vec e_{4\e}.\vec e_5)   \\
 0      &  -b_5e^{-\a}     & 0 &  (\vec e_{5\e}.\vec e_4)     & 0
\end{pmatrix}.
\label{p8} 
} 
One can also find the equations for the coefficients $a_i$ and $b_j$ 
\al{ 
& a_{4\e}=a_5(\vec e_5.\vec e_{4\e}), \quad
a_{5\e}=a_4(\vec e_4.\vec e_{5\e}) \\
& b_{4\x}=b_5(\vec e_5.\vec e_{4\x}), \quad b_{5\x}=b_4(\vec
e_4.\vec e_{5\x}). 
} 
Having in mind that $(\vec e_4.\vec e_{5\e})=-(\vec e_{4\e}.\vec e_5)$ 
and analogously $(\vec e_4.\vec e_{5\x})=-(\vec e_{4\x}.\vec e_5)$, one find 
\eq{
\p_\e[a_4^2+a_5^2]=0, \quad \p_\x[b_4^2+b_5^2]=0, } or, \al{
& a_4=P(\e)\cos\d(\x,\e), \quad  a_5=P(\e)\sin\d(\x,\e), \label{p-sol1}\\
& b_4=Q(\x)\cos\gamma(\x,\e), \quad  b_5=Q(\x)\sin\gamma(\x,\e).
\label{p-sol2} 
} 
The compatibility condition for the equations (\ref{p5-a},\ref{p5-b}) is the well-known zero curvature condition
\eq{ 
\p_\e \mathbf{A}-\p_\x \mathbf{B}+[\mathbf{A},\mathbf{B}]=0.
\label{comp-ads} 
} 
According to \eqref{p-sol1} and \eqref{p-sol2}
\eq{ 
(\vec e_4.\vec e_{5\x})=\gamma_\x, \quad (\vec e_4.\vec
e_{5\e})=\d_\e, 
} 
which combined with the compatibility condition \eqref{compatib}, $\p_\e
\mathbf{A}-\p_\x\mathbf{B}+[\mathbf{A},\mathbf{B}]=0$, gives the
equations for the dynamical variables 
\al{
& \a_{\x\e}(\x,\e)-e^{\a(\x,\e)}-\lb(a_4b_4+a_5b_5\rb)e^{-\a(\x,\e)}=0 \label{p9} \\
& \b_{\x\e}(\x,\e)-\lb(a_4b_5-a_5b_4\rb)e^{-\a(\x,\e)}=0, \quad
\b(\x,\e)=\gamma-\d. \label{p10} 
} 
One can use the explicit expressions for $a_i$ and $b_i$ to obtain 
\al{
& \a_{\x\e}(\x,\e)-e^{\a(\x,\e)}-Q(\x)P(\e)e^{-\a(\x,\e)}\cos\b=0 \label{p9a} \\
& \b_{\x\e}(\x,\e)-Q(\x)P(\e)e^{-\a(\x,\e)}\sin\b=0. \label{p10a}
} 
Let us make the transformations 
\eq{ 
\a(\x,\e)=\hat
\a(x,y)+\log[F(\x)G(\e)]. \label{p11} 
} 
Choosing 
\al{
& \frac{d\,x}{d\x}=F(\x), \quad \frac{d\,y}{d\e}=G(\e),\notag \\
& F^2(\x)G^2(\e)=-Q(\x)P(\e) \label{p12} 
} 
we find 
\al{
\hat\a_{xy}(x,y)-e^{\hat\a(x,y)}+e^{-\hat\a(x,y)}\cos\b(x,y)=0 \label{p13} \\
\b_{xy}(x,y)+e^{-\hat\a(x,y)}\sin\b(x,y)=0. \label{p14} 
}

One must note that the derivation of the above result relies on
the Virasoro constraints \eqref{vir-p} defining  $\vec Y_\x$ and
$\vec Y_\e$ as null-vectors. This, however, is not the general
case. We proceed with the more general case in the next paragraph.


\paragraph{Reduction of $AdS_4$ part of string sigma model.} \ \\

Let us consider the string sigma model on $AdS_4\times\cp$. As we
already discussed above, this case is different because although
the total energy-momentum is vanishing, the energy-momentum of
each of the two parts in the product space is a non-vanishing
constant \eqref{Vir-tot}. In this case we will shortly present the
method of \cite{Grigoriev:2007bu,Grigoriev:2008jq,Bakas:1995bm}
applied to this concrete case.

We start with the parametrization of the Lax connection. The
linear problem associated with our sigma model is defined by 
\al{
& \p_\x\f(\x,\e,\z)=\mathbf{L}^a(\x,\e,\z)\f(\x,\e,\z) \notag \\
& \p_\e\f(\x,\e,\z)=\mathbf{M}^a(\x,\e,\z)\f(\x,\e,\z).
\label{cp-1} 
} 
The general dependence on $\z$ is as follows 
\eq{
 \mathbf{L}^a=\z \mathbf{A}^a+\mathbf{C}^a , \quad \mathbf{M}^a=\z^{-1}\mathbf{B}^a+\mathbf{D}^a.
\label{cp-2} 
} 
The consistency condition of the above defined linear problem reads off 
\eq{ 
\p_\e \mathbf{L}^a-\p_\x
\mathbf{M}^a+[\mathbf{L}^a,\mathbf{M}^a]=0 \label{cp-3} 
} 
and it splits into 
\al{
& \p_\e \mathbf{A}^a+[\mathbf{A}^a,\mathbf{D}^a]=0 \label{cp-4a} \\
& \p_\x \mathbf{B}^a+[\mathbf{B}^a,\mathbf{C}^a]=0 \label{cp-4b} \\
& \p_\e \mathbf{C}^a-\p_\x
\mathbf{D}^a+[\mathbf{A}^a,\mathbf{B}^a]+
[\mathbf{C}^a,\mathbf{D}^a]=0. \label{cp-4c} 
} 
The general form of the matrices $\mathbf{A}^a$ and $\mathbf{B}^a$ is 
\eq{
\mathbf{A}^a=
\begin{pmatrix}
 0 & \bar\psi_1 & \bar\psi_2 & \bar\psi_3 \\
-\psi_1 & 0 & 0 & 0 \\
-\psi_2 & 0 & 0 & 0 \\
-\psi_3 & 0 & 0 & 0
\end{pmatrix}
\label{cp-5} 
} 
and 
\eq{ 
\mathbf{B}^a=\kappa
\begin{pmatrix}
 0 &  Y_1 & Y_2 &  Y_3 &  Y_4\\
- Y_1 & 0 & 0 & 0 & 0\\
- Y_2 & 0 & 0 & 0 & 0\\
- Y_3 & 0 & 0 & 0 & 0\\
- Y_4 & 0 & 0 & 0 & 0
\end{pmatrix}.
\label{cp-6} 
} 
Here $Y_i$ satisfy the relation 
\eq{
Y_1^2-\sum\limits_{k=2}^2Y_k^2=1. \label{rel-ads-f} 
} 
With the help of certain gauge transformations one can make
$\mathbf{D}^a=0$. Then the matrices $\mathbf{L}^a$ and $\mathbf{M}^a$ can be brought into the form 
\eq{ 
\mathbf{L}^a=\z
\mathbf{A}^a+\mathbf{C}^a, \quad \mathbf{M}^a=\z^{-1}\mathbf{B}^a,
\label{cp-7} 
} 
where 
\eq{ 
\mathbf{A}^a=\kappa
\begin{pmatrix}
  0 &-1 & 1 & 0 & 0 \\
 -1 & 0 & 0 & 0 & 0 \\
  1 & 0 & 0 & 0 & 0 \\
  0 & 0 & 0 & 0 & 0 \\
  0 & 0 & 0 & 0 & 0
\end{pmatrix}.
\label{cp-8} 
} 
The matrix $\mathbf{C}^a$ belongs to the centralizer of $SO(2,5)$ and has the form 
\eq{ 
\mathbf{C}^a=
\begin{pmatrix}
  0 & 0 & 0 & 0 & 0\\
  0  & 0 & -c_2 & - c_3 &  c_4\\
  0  & c_2 & 0 & 0 & 0\\
  0  & c_3 & 0 & 0 & 0 \\
  0  & c_4 & 0 & 0 & 0
\end{pmatrix}.
\label{cp-9} 
} 
Eliminating $c_k$, we end up with the equations
\eq{ 
\p_\x\dfrac{\p_\e
Y_k}{\sqrt{1+\sum_{l=2}^4Y_l^2}}=-\kappa^2Y_k,\quad k=2,3,4.
\label{eom-ads-f} 
} 
One can use the following convenient parametrization \cite{Grigoriev:2007bu,Grigoriev:2008jq} 
\eq{
Y_1=\cosh 2\f, \quad Y_k=r_k\,\sinh 2\f, \quad
\sum\limits_{j=2}^4r_l^2=1, \quad k=2,3,4. \label{param-ads-f} 
}
Substituting into \eqref{eom-ads-f} and using \eqref{rel-ads-f} we find 
\al{
& \p_\x\p_\e \f-\frac{1}{2}\tanh 2\f\sum\limits_{l=2}^4\p_\e r_l\p_\x r_l+\frac{\kappa^2}{2}\sinh 2\f=0 \\
& \p_\x\p_\e r_l+(\sum\limits_{k=2}^4\p_\x r_k\p_\e r_k)r_l+
\frac{2}{\sinh 2\f}\lb(\cosh 2\f\p_\e\f\p_\x r_l+\frac{1}{\cosh
2\f}\p_\x\f\p_\e r_l\rb)=0, \quad l=2,3,4. \label{eqs-ads-f} 
}

We must note that to further reduce the degrees of freedom one may
need to fix the residual conformal symmetry. For instance, in the
lower dimensional $AdS_3$ case such a gauge fixing relates the
above approach to the sinh-Gordon equation for a single dynamical
field \cite{Grigoriev:2008jq}


\subsection{Reduction of $\cp$}

In this Section we investigate the reduction of the string sigma
model on $\cp$.

\paragraph{General remarks}\ \\

Let us briefly review the basic properties of $\mathbb{CP}^{n}$.
The most convenient way to define
 n-dimensional complex projective space $\mathbb {CP}^n$ is as the family of one-dimensional subspaces
in $\mathbb{C}^{n+1}$, i.e. this is the quotient
$\mathbb{C}^{n+1}/(\mathbb{C}\setminus\{0\})$. The equivalence
relation is defined as 
\eq{ \a Z_1:\cdots:\a
Z_{n+1}=Z:\cdots:Z_{n+1}.\notag 
} 
The space $\mathbb {CP}^n$ itself is covered by patches
$U_i:\{Z_1:\cdots:Z_{n+1}\in\mathbb{CP}^n\,|\,Z_i\neq 0\}, \,\,
i=1,\cdots,n+1$. One can see that each patch $U_i$ is isomorphic
to $\mathbb {CP}^n$, where the isomorphism is defined by
$W_j^{(i)}=Z_j/Z_i,\,\, j\neq i$. One can choose local coordnates
$W=(W_1,W_2,\cdots,W_n)^t\in\mathbb{C}^{n+1}$ with $W_j\equiv
W_j^{(n+1)}$. The Fubini-Study metric then is given by the line
element 
\eq{ ds^2=\frac{(1+|W|^2)|dW|^2-|W^\dagger dW|^2}
{(1+|W|^2)^2}. \label{cp-FSmetric} 
}

One can think of $\mathbb {CP}^n$ as the homogeneous space
$\mathbb {CP}^n=U(n+1)/(U(n)\times U(1))$. The $u(n+1)$ Lie
algebra $\mathfrak{f}$ can be realized as anti-hermitian matrices
and splits into two parts: $\mathfrak{p}=u(n)\oplus u(1)$ and its
orthogonal completion $\mathfrak{cp}(n)$ with respect to the
$U(n+1)$ Killing form 
\al{
& \mathfrak{p}=u(n)\oplus u(1)=\{iM\in u(n+1)\,|\,[\Gamma,M]=0\} \notag \\
& \mathfrak{cp}(n)=\{iM\in u(n+1)\,|\,\{\Gamma,M\}=0\}, 
} 
where $M$ is traceless and hermitian and
$$ \Gamma=\begin{pmatrix}-1 & \\ & \mathbf{1}_n \end{pmatrix}.$$
Using diagonal embedding of $\mathfrak{p}$, $\mathbb {CP}^n$ can
be though as an orbit in the coset with a  generator of
$\mathfrak{cp}(n)$ part, $\mathbf{B}$ then is given by 
\eq{
\mathbf{B}=\begin{pmatrix} & W^\dagger \\ -W \end{pmatrix}. 
} 
Then one can write schematically 
\eq{
 \mathfrak{f}=\mathfrak{p}\oplus \mathfrak{cp}, \quad [\mathfrak{p},\mathfrak{p}]\subset \mathfrak{p}, \quad
[\mathfrak{p},\mathfrak{cp}]\subset \mathfrak{cp}, \quad
[\mathfrak{cp},\mathfrak{cp}]\subset \mathfrak{cp}. 
}

The linear problem associated with our sigma model is defined by
\al{
& \p_\x\f(\x,\e,\z)=\mathbf{L}(\x,\e,\z)\f(\x,\e,\z) \notag \\
& \p_\e\f(\x,\e,\z)=\mathbf{M}(\x,\e,\z)\f(\x,\e,\z).
\label{cp-1g} } The general dependence on $\z$ is as follows \eq{
 \mathbf{L}=\z \mathbf{A}+\mathbf{C} , \quad \mathbf{M}=\z^{-1}\mathbf{B}+\mathbf{D}.
\label{cp-2g} } The consistency condition of the above defined
linear problem reads off \eq{ \p_\e \mathbf{L}-\p_\x
\mathbf{M}+[\mathbf{L},\mathbf{M}]=0 \label{cp-3g} } and it splits
into \al{
& \p_\e \mathbf{A}+[\mathbf{A},\mathbf{D}]=0 \label{cp-4ag} \\
& \p_\x \mathbf{B}+[\mathbf{B},\mathbf{C}]=0 \label{cp-4bg} \\
& \p_\e \mathbf{C}-\p_\x
\mathbf{D}+[\mathbf{A},\mathbf{B}]+[\mathbf{C},\mathbf{D}]=0.
\label{cp-4cg} } 
The general form of the matrices \textbf{A} and
\textbf{B} is \eq{ \mathbf{A}=
\begin{pmatrix}
 0 & \bar\psi_1 & \bar\psi_2 & \bar\psi_3 \\
-\psi_1 & 0 & 0 & 0 \\
-\psi_2 & 0 & 0 & 0 \\
-\psi_3 & 0 & 0 & 0
\end{pmatrix}
\label{cp-5g} } and \eq{ \mathbf{B}=
\begin{pmatrix}
 0 & \bar W_1 & \bar W_2 & \bar W_3 \\
- W_1 & 0 & 0 & 0 \\
- W_2 & 0 & 0 & 0 \\
- W_3 & 0 & 0 & 0
\end{pmatrix}.
\label{cp-6g} 
} 
By making use of the gauge transformations one can
make \textbf{D}=0 and bring the matrices \textbf{L} and \textbf{M}
into the form 
\eq{ \mathbf{L}=\z \mathbf{A}+\mathbf{C}, \quad
\mathbf{M}=\z^{-1}\mathbf{B}, \label{cp-7g} 
} 
where 
\eq{
\mathbf{A}=
\begin{pmatrix}
 0 &-1 & 0 & 0 \\
 1 & 0 & 0 & 0  \\
 0 & 0 & 0 & 0  \\
 0 & 0 & 0 & 0
\end{pmatrix}.
\label{cp-8g} 
} 
\eq{ 
\mathbf{C}=
\begin{pmatrix}
 c_1 & 0 & 0 & 0 \\
  0  & -c_1 & -\bar c_2 & -\bar c_3 \\
  0  & c_2 & 0 & 0 \\
  0  & c_3 & 0 & 0
\end{pmatrix}.
\label{cp-9g} 
} 
Since we are dealing with $\cp$ sigma model, one can always normalize the fields as 
\eq{ 
| W_1|^2+| W_2|^2+| W_3|^2=1.\label{cp-norm} 
} 
Using this normalization, the entries
of the matrix \textbf{C} are given by 
\al{
& c_1=\frac{W_1\p_\x\bar W_1+\p_\x W_2.\bar W_2+\p_\x W_3.\bar W_3}{3| W_1|^2-1}, \label{cp-10g} \\
& c_k=\frac{\p_\x W_k}{ W_1}+\frac{ W_k}{ W_1}.c_1, \quad k=2,3.
\label{cp-11g} 
} 
The condition \eqref{cp-4cg} gives 
\al{
& \p_\e c_1+\bar W_1- W_1=0 \notag \\
& \p_\e c_k+ W_k=0, \quad k=2,3. \label{cp-12g} 
}

The reduced system involves 5 independent real scalar fields.

\paragraph{Equations of motion for the reduced model}\ \\

We will use the following parametrization 
\eq{
W_i=r_ie^{i\f_i}, \quad i=1,2,3, \label{cp-13} 
} 
which implies
\eq{
 r_1^2+r_2^2+r_3^2=1. 
} 
The matrix \textbf{B} takes the form
\eq{ 
\mathbf{B}=
\begin{pmatrix}
 0 & r_1e^{-i\f_1} & r_2e^{-i\f_2} & r_3e^{-i\f_3} \\
 -r_1e^{i\f_1}& 0 & 0 & 0 \\
-r_2e^{i\f_2} & 0 & 0 & 0 \\
- r_3e^{i\f_3}& 0 & 0 & 0
\end{pmatrix}.
\label{cp-14} } The entries of the matrix \textbf{C} are 
\al{
& c_1=\frac{r_2^2\p_\x\f_{2}+r_3^2\p_\x\f_{3}-r_1^2\p_\x\f_{1}}{3r_1^2-1}:=\frac{i}{2}\p_\x\b \label{cp-15a} \\
& c_k=\frac{e^{i(\f_k-\f_1)}}{r_1}\lb[\p_\x
r_{k}+ir_k(\p_\x\f_{k}+\frac{1}{2}\p_\x\b)\rb], \quad k=2,3.
\label{cp-15b} } Let us determine one of the angles, say the angle
$\f_2$, through the other angles. Using \eq{
r_2^2\p_\x\f_2+r_3^2\p_\x\f_3-r_1^2\p_\x\f_1=\frac{1}{2}(3r_1^2-1)\p_\x\b
} 
and the equations (\ref{cp-norm}-\ref{cp-12g}) one finds that
\eq{
r_2\p_\xi(\f_2+\frac{1}{2}\b)=\frac{r_1^2}{r_2}\p_\xi\f_1-\frac{r_3^2}{r_2}\p_\xi
+\frac{1}{2r_2}(2 r_1^2-r_3^2)\p_\xi\b, \label{cp-16} } and \eq{
\p_\e(\f_2-\f_1)=-\frac{1}{r_2^2}\lb[\p_\e\b+(r_1^2+r_2^1)\p_\e\f_1+
r_3^2\p_\e\f_3\rb]. \label{cp-17} } Then \al{ &
c_2=e^{i(\f_2-\f_1)}\lb[\frac{\p_\xi r_2}{r_1}+i
\lb(\frac{r_1^2\p_\xi\f_1-r_3^2\p_\xi\f_3}{r_1r_2}+
\frac{2r_1^2-r_3^2}{2r_1r_2}\p_\xi\b\rb)\rb], \label{cp-18} \\
& c_3=e^{i(\f_3-\f_1)}\lb[\frac{\p_\xi r_3}{r_1}+i
\frac{r_3}{r_1}\p_\xi(\f_3+\frac{1}{2}\b)\rb]. \label{cp-19} 
}
After long and tedious, but straightforward calculations one finds
the equation of motion of the dynamical variables of the system.
They are found to be 
\ml{
 \Box r_2+\frac{\p_\xi r_2\p_\e r_2}{r_2}-\frac{\p_\e(r_1r_2)\p_\xi r_2}{r_1r_2} \\+
\frac{(\p_\e\b+(r_1^2+r_2^2)\p_\e\f_1+r_3^2\p_\e\f_3)(r_1^2\p_\xi\f_1-r_3^2\p_\xi(\f_3+\frac{1}{2}\b)+
r_1^2\p_\xi\b)}{r_2^3}\\
+r_1r_2\cos\f_1=0, \label{cp-20} } \ml{
 \Box\f_1+\frac{r_3[\p_\xi r_3\p_\e(\f_3-\f_1)-\p_\e r_3\p_\xi(\f_3+\frac{1}{2}\b)]}{r_1^2}
+\frac{r_3^2\p_e r_1\p_\xi(\f_3+\frac{1}{2}\b)}{r_1^3}\\
+\frac{2\p_\e r_1\p_\xi(\f_1+\b)}{r_1}-\frac{\p_\xi r_2[\p_\e(\f_1+\b)+r_3^2\p_\e(\f_3-\f_1)]}{r_1^2r_2}\\
+\frac{2(2r_1^2+r_3^2)+r_1r_2}{r_1}\sin\f_1=0 } \al{
& \Box\b-4r_1\sin\f_1=0, \label{cp-22}\\
& \Box r_3-\frac{\p_\xi r_3\p_\e
r_1}{r_1}-r_3\p_\e(\f_3-\f_1)\p_\xi(\f_3+\frac{1}{2}\b)+
r_1r_3\cos\f_1=0, \label{cp-23}\\
& \Box\f_3+\frac{\p_\e r_3\p_\xi(\f_3+\frac{1}{2}\b)+\p_\xi
r_3\p_\e(\f_3-\f_1)}{r_3} -\frac{\p_\e
r_1\p_\xi(\f_3+\frac{1}{2}\b)}{r_1}+3r_1\sin\f_1=0 \label{cp-24} 
}
Note that $r_1^2+r_2^2+r_3^2=1$ so that the number of dynamical
variables is five.

To reduce the $\cp$ system to the ${\mathbb {CP}}^2$ case one has
(carefully) to set $r_3=\f_3=0$. It is easy then (using the
parametrization $r_1=\cos\a, r_2=\sin\a$) to see that the
equations reduce to the known ones.


\sect{Conclusions}

This study is inspired by the recent breakthrough in our
understanding of membrane dynamics and its application to
$AdS_4/CFT_3$ correspondence \cite{ABJM}. The hopes are that this
is another example of exact duality between gauge theory and
strings/M-theory. The wide range of possible applications make the
subject even more attractive. This strongly motivates the
intensive research on various aspects of string theory on
$AdS_4\times\cp$ background. The experience from the well studied
$\axs$ case teaches us that the techniques coming from integrable
systems and the study of integrable structures play an important
role.

In this note we initiated a more detailed analysis of the
reduction of string sigma model on the $AdS_4\times\cp$
background. We presented a short analysis of the Pohlmeyer
reduction of the $AdS_4$ part of the sigma model action and its
extension to the string sigma models action. After we performed a
reduction of the $\cp$ part of the Polyakov string action. As a
result we found the equations of motion for the dynamical degrees
of freedom of the reduced model. Certainly this study is far from
complete, nevertheless it seems to be good basis to proceed
further. For instance, it would be interesting to use B\"{a}cklund
transformations (or dressing method) to generate nontrivial
solutions. Even restricted to some subspaces such investigations
would be very useful.

\textit{Note added.} After this paper was sent to the Arxiv another interesting and related to our study
paper appeared \cite{Miramontes:2008wt}. It presents a systematic study of Pohlmeyer reduction with emphasis
to the Lagrangian formulation. Although there is a partial overlap, in general the results in the two papers 
are complimentary and may be useful in further study of AdS/CFT correspondence.


\section*{Acknowledgements}
This work was supported in part  by NSFB VU-F-201/06 and by the
Austrian Research Fund FWF (grant \# P19051-N16). The author
acknowledges  warm hospitality at the Institute for Theoretical
Physics, Vienna University of Technology where this project was
mainly carried out. Many thanks to Max Kreuzer and his group for
fruitful and stimulating atmosphere. I thank N. Bobev for
critically reading the manuscript.



\section*{Appendix: Reduction of $AdS_4\times\cp$ sigma model}

\def\theequation{A.\arabic{equation}}
\setcounter{equation}{0}
\begin{appendix}

The two seemingly unrelated parts of $AdS_4\times\cp$ sigma model
actually are related through the Virasoro constraints 
\eq{
T_{\pm\pm}(AdS_4)+T_{\pm\pm}(\cp)=0. \label{vir-gen} 
} 
The reduction above then have to be modified. Actually one can show
(see for instance \cite{Dimov:2007ey,Ahn:2008hj}) that 
\eq{
T_{\pm\pm}(\cp)=\m^2=-T_{\pm\pm}(AdS_4), 
\label{vir-split} 
} 
which causes small but important modifications in the above reduction.
One must note that there are cases when the string dynamics is
confined in one part of the geometry, $S^5$ or $\cp$. In this case
we have $T_{\pm\pm}(\cp)=0=T_{\pm\pm}(AdS_4)$ and these cases
should be derivable taking the limit $\m\rightarrow 0$ and
therefore these cases are covered by the analysis above.

We find for the matrix \textbf{A} the expression 
\eq{ 
\mathbf{A}=\begin{pmatrix}
 0      & 1      & 0           & 0      & 0    \\
 -\m^2      & \frac{\a_\x e^{\a}}{\Upsilon}  & \frac{\a_\x \m^2}{\Upsilon}           & a_4    & a_5   \\
 e^\a   & 0      & 0           & 0      & 0       \\
 0      & -\frac{\m^2a_4 e^{-\a}}{\Upsilon}       & -\frac{a_4}{\Upsilon} & 0      & (\vec e_{4\x}.\vec e_5)   \\
 0      & -\frac{\m^2a_5 e^{-\a}}{\Upsilon}       & -\frac{a_5}{\Upsilon} &  (\vec e_{5\x}.\vec e_4)     & 0
\end{pmatrix}
\label{sigma-1} 
} 
where $\Upsilon=e^\a-\m^4 e^{-\a}$. Analogously one can find for the matrix \textbf{B} 
\eq{ 
\mathbf{B}=\begin{pmatrix}
 0      & 0      & 1           & 0      & 0    \\
 e^\a   & 0      & 0           & 0      & 0   \\
 -\m^2      & \frac{\m^2\a_\e}{\Upsilon}      &  \frac{\a_\e e^{\a}}{\Upsilon}      & b_4    & b_5       \\
 0      & -\frac{b_4}{\Upsilon}    & -\frac{\m^2b_4 e^{-\a}}{\Upsilon} & 0      & (\vec e_{4\e}.\vec e_5)   \\
 0      & -\frac{b_5}{\Upsilon}    & -\frac{\m^2b_5 e^{-\a}}{\Upsilon} &  (\vec e_{5\e}.\vec e_4)     & 0
\end{pmatrix}
\label{sigma-2} }

The compatibility condition \eqref{comp-ads} gives the following
equations for the entries of \textbf{A} and \textbf{B}: 
\al{ 
& \Box \a-\frac{\m^4\a_\x\a_\e e^{-\a}}{e^\a-\m^4 e^{-\a}}-e^\a+\m^4
e^{-\a}-(a_4b_4+a_5b_5)e^{-\a}=0
\label{sigma-3} \\
& \p_\e a_4+\frac{\m^2\a_\x}{e^\a-\m^4 e^{-\a}}b_4+a_5(\vec
a_{5\e}.\vec a_4)=0, \quad
\p_\e a_5+\frac{\m^2\a_\x}{e^\a-\m^4 e^{-\a}}b_5+a_4(\vec a_{4\e}.\vec a_5)=0 \label{sigma-4} \\
& \p_\x b_4+\frac{\m^2\a_\e}{e^\a-\m^4 e^{-\a}}a_4+b_5(\vec
b_{5\x}.\vec b_4)=0, \quad \p_\x b_5+\frac{\m^2\a_\e}{e^\a-\m^4
e^{-\a}}a_5+b_4(\vec b_{4\x}.\vec b_5)=0. \label{sigma-5} 
} 
From these one can derive
 \al{
& \p_\e(a_4^2+a_5^2)=-(a_4b_4+a_5b_5)\frac{\m^2\a_\x}{\Upsilon} \label{sigma-6} \\
& \p_\x(b_4^2+b_5^2)=-(a_4b_4+a_5b_5)\frac{\m^2\a_\e}{\Upsilon},
\label{sigma-7} 
} 
or, 
\eq{
\frac{\m^4\a_\x\a_\e}{\Upsilon^2}=\frac{\p_\x(b_4^2+b_5^2)\p_\e(a_4^2+a_5^2)}{(a_4b_4+a_5b_5)^2}.
} Then one can write \eq{ \Box
\a-\frac{\p_\x(b_4^2+b_5^2)\p_\e(a_4^2+a_5^2)}{(a_4b_4+a_5b_5)^2}e^{-\a}
(e^\a-\m^4 e^{-\a})-(e^\a-\m^4 e^{-\a})-(a_4b_4+a_5b_5)e^{-\a}=0,
} or \eq{ \Box
\a-\frac{\p_\x(b_4^2+b_5^2)\p_\e(a_4^2+a_5^2)}{(a_4b_4+a_5b_5)^2}
(1-\m^4 e^{-2\a})-(e^\a-\m^4 e^{-\a})-(a_4b_4+a_5b_5)e^{-\a}=0. 
}
When $\m\rightarrow 0$, the r.h.s. of \eqref{sigma-6} and
\eqref{sigma-7} vanishes and the equation for $\a(\x,\e)$ reduces
to that in (\ref{p13}-\ref{p14}). Note that the vanishing of the
r.h.s. of  \eqref{sigma-6} and \eqref{sigma-7} means $\p_\e a_k=0$
and $\p_\x b_k=0$ (k=1,2) which is related to the fixing of the
conformal symmetry.

\end{appendix}



\end{document}